# Tag Clouds for Software Documents Visualization

Ra'Fat Al-Msie'deen#

# Department of IT, Faculty of Information Technology, Mutah University, Mutah 61710, Karak, Jordan
E-mail: rafatalmsiedeen@mutah.edu.jo

*Abstract*— Legacy software documents are hard to understand and visualize. The tag cloud technique helps software developers to visualize the contents of software documents. A tag cloud is a well-known and simple visualization technique. This paper proposes a new method to visualize software documents, using a tag cloud. In this paper, tags visualize in the cloud based on their frequency in an alphabetical order. The most important tags are displayed with a larger font size. The originality of this method is that it visualizes the contents of JavaDoc as a tag cloud. To validate the JavaDocCloud method, it was applied to NanoXML case study, the results of these experiments display the most common and uncommon tags used in the software documents.

## I. INTRODUCTION

Nowadays, tag clouds are widely used in several domains. For instance, tag clouds have been used in the software engineering domain for software source code and information visualisation [1-4]. This paper suggests an original approach called JavaDocCloud to visualize software documents as a tag cloud. This paper considers only the Javadoc document. The Javadoc is a software document written by the software developer to summarize software source code [5]. The tag cloud is a visualization technique [6], while the tag is a single word. The main idea of the tag cloud is to visualize the content of document as a single tag cloud. The tag cloud uses font size to display the tag frequency in the document.

Most current methods are designed to identify tag clouds from textual documents. In the literature, there is no approach identifies tag cloud from software documents such as Javadoc. There are very few existing methods which use a tag cloud visualization technique with software source code. Emerson et al. use tag clouds to visualize class names and method names [1, 2]. Cottrell et al. suggest an approach to visualize software methods through tag cloud [7]. Al-msie'deen [4] offers the Iconic method to visualize software source code as a tag cloud.

JavaDocCloud approach aims to help software developer to understand the documents of legacy software system. Tag cloud helps software developers to discover the most frequent words used in the Javadoc. JavaDocCloud approach accepts as input the software document (Javadoc). Then, the approach extracts all words from Javadoc. Next, the approach stems the words into their roots or tags. After that, the approach assigns weight to each tag based on its frequency in the document. Finally, JavaDocCloud identifies tag cloud as output. This approach is based on the previous work called Iconic [4].

The JavaDocCloud approach is detailed in the remainder of this paper as follows. Section II discusses the related work. Section III presents Javadoc tag cloud process step by step. Section IV describes the experiments that were conducted to validate JavaDocCloud proposal, while section V concludes and provides perspectives for this work.

## II. RELATED WORK

This section presents the related work relevant to JavaDocCloud contributions. It also provides a concise overview of the different approaches and shows the need to propose JavaDocCloud approach. Anslow et al. [8] use a tag cloud to visualize the structure of Java class names. The extracted tag clouds have exposed the most frequently used words used in Java class names. Cottrell et al. [7] propose an approach to visualize software methods via tag cloud. Eclipse plugin Sourcecloud [9] creates a tag cloud visualization of the text within software code. The authors of [10] used the tag cloud to help name the extracted features based on the most frequent word in the identified blocks. In [4], the author used a tag cloud to visualize software identifier names (i.e. packages, classes, attributes and methods). In software engineering domain, most existing approaches are designed to extract tag clouds from software source code. Regarding software documents there is no approach that visualizes these documents as a tag cloud. The concise overview of the existing approaches shows the need to propose an approach to visualize the software documents by using the tag cloud.

## III. JAVADOC TAG CLOUD PROCESS

This section presents the key ideas and principles used in the Javadoc Cloud approach and, at last, it describes the Javadoc tag cloud process step by step.

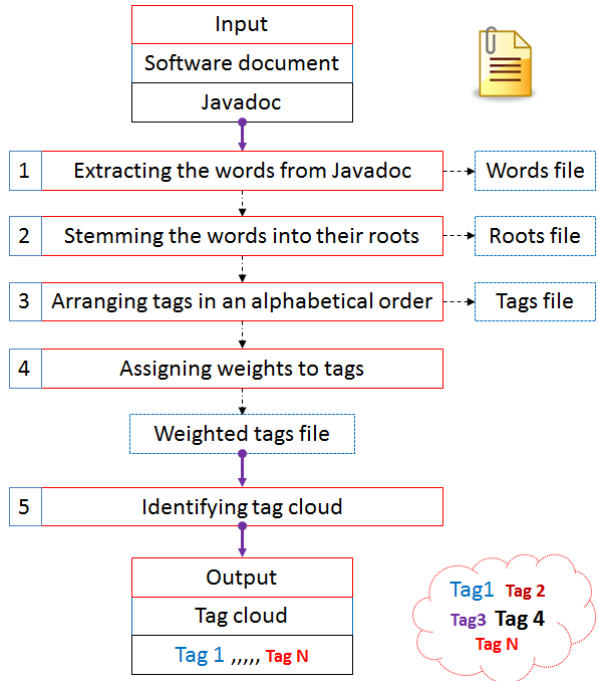

Fig. 1 The Javadoc tag cloud process

An overview of the Javadoc tag cloud process is shown in Figure 1. The input is the legacy Javadoc documents obtained from software documents. The output is the tag cloud. As an illustrative example, this paper considers the Javadoc of the ArgoUML [11] case study. The JavaDocCloud approach identifies the Javadoc tag cloud in five steps as detailed in the following.

### A. Extracting the Words from Javadoc

The first step of the Javadoc tag cloud process is the extraction of words from Javadoc documents. The goal of this step is to mine all the words from the Javadoc documents. The inputs of this step are a set of Javadoc documents. The output of this step is a document containing all the words from Javadoc documents. In this step, the punctuation marks and numbers are removed. The approach splits the words of document into a set of words by using the camel-case splitting algorithm [12, 13]. The camel-case splitting algorithm splits words based on capital letters. For example: skipWhitespace is split into a skip and whitespace.

### B. Stemming the Words into Their Roots

The second step of the Javadoc tag cloud process is the word stemming by using WordNet [14, 15]. Word stemming means dropping a word to its word stem or root. For example, the word "combined" has the word root "combine". The stemmer takes as input the original word and produces as output the word base or tag. The input of this step is the words file which is the output of the previous step. The output of this step is the tags file.

### C. Arranging Tags in An Alphabetical Order

The third step of the Javadoc tag cloud process is arranging tags in an alphabetical order. This step accepts as input the tags file from the previous step and creates as output a tags file which contains all tags in an alphabetical order. For example, a tags file has the following tags: element, charlie, beta and alpha. In order to display these tags in the cloud there is a need to store these tags in an alphabetical order. Thus, the output tags file contains the same tags in the following order: alpha, beta and charlie, element.

### D. Assigning Weights to Tags

The fourth step of the Javadoc tag cloud process is a tag weighting. This step accepts as input the tags file from the previous step and creates as output a tags file which contains the frequency of each tag across all Javadoc documents. The font size of each tag in the tag cloud gives an indication about the tag importance. In this step, a weight is given to each tag. Where a weight is given to tag based on its frequency across all software documents. In tag cloud, larger font sizes assigning to the more frequent tags. For example, in the Javadoc of legacy software the message tag was occurred 5 times across all documents, so the given weight of this tag is 5.

### E. Identifying Tag Cloud

The last step of the Javadoc tag cloud process is identifying the tag cloud. This step accepts the tags file and builds the tag cloud as output. Tag cloud represents all tags extracted from software documents. The tags importance in the cloud can obtain easily through tag font sizes. In this paper, the weight of the tag is determined by the font-size only. The use of color is random, where it is not mapped to any conditions. The mined tag cloud allows software developers to comprehend what are the most common tags as well as the uncommon tags.

As an example, the approach uses the Javadoc for the main class of Argo UML [16] software which belongs to org.argouml.application package. Table I shows the Javadoc for the main class of ArgoUML software.

TABLE I
JAVADOC FOR MAIN CLASS OF ARGOUML SOFTWARE

| Class Main Summary | |
|---|---|
| Package org.argouml.application | |
| Here it all starts ... | |
| **Field Summary** | |
| static String | DEFAULT_LOGGING_CONFIGURATION: The location of the default logging configuration (.lcf) file. |
| private static Logger | LOG Logger. |
| private static Vector | postLoadActions |
| **Constructor Summary** | |
| Main() | |
| **Method Summary** | |
| static void | addPostLoadAction(Runnable r): Add an element to the PostLoadActions list. |
| private static void | checkHostsFile(): Check that we can get the InetAddress for localhost. |
| private static void | checkJVMVersion(): Checks the JVM Version. |
| private static SplashScreen | initializeGUI(boolean doSplash, String the Theme): Do a part of the initialization that is very much GUI-stuff. |
| static void | main(String[] args): The main entry point of ArgoUML. |
| static void | performCommands(List list): Perform a list of commands that were given on the command line. |
| private static void | printUsage(): Prints the usage message. |
| private static URL | projectUrl(String projectName, URL urlToOpen): Calculates the URL for the given project name. |

Figure 2 shows the extracted tag cloud from Javadoc of main class in Table I. In Figure 2, tag cloud shows the most common tag (e.g., static and the) and the uncommon tags (*e.g.,* args, use, we, much and part).

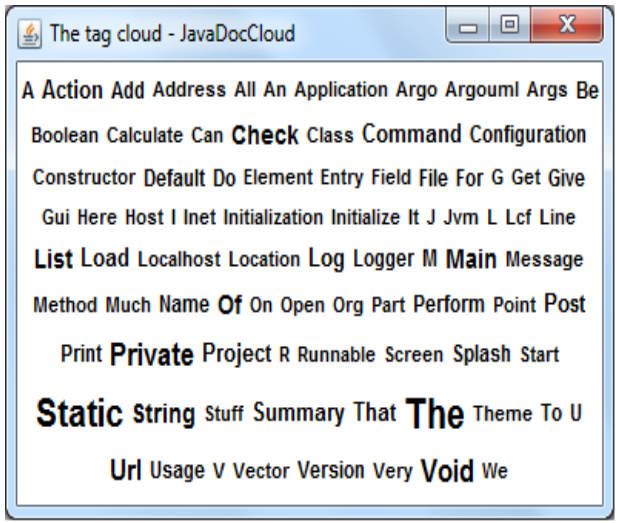

Fig. 2 The tag cloud of main class

Figure 3 shows the tag cloud for the main class of ArgoUML software system. In this tag cloud, the number of tag frequency appears beside each tag.

## IV. EXPERIMENTATION

This section presents the experiment that conducted to validate the JavaDocCloud approach. It also shortly presents the NanoXML case study and its results and, at last, it presents the threats to validity of JavaDocCloud approach.

The JavaDocCloud approach has been tested on the NanoXML [17] case study. The NanoXML software is a Java program for parsing XML file. The approach prototype is produced to extract tag cloud from software Javadoc. More information about prototype is available at JavaDocCloud web page [18]. Figure 4 shows the extracted tag cloud from JavaDoc of NanoXML software. The JavaDocCloud approach needs to 1988 ms to create the tag cloud from the JavaDoc of NanoXML software.

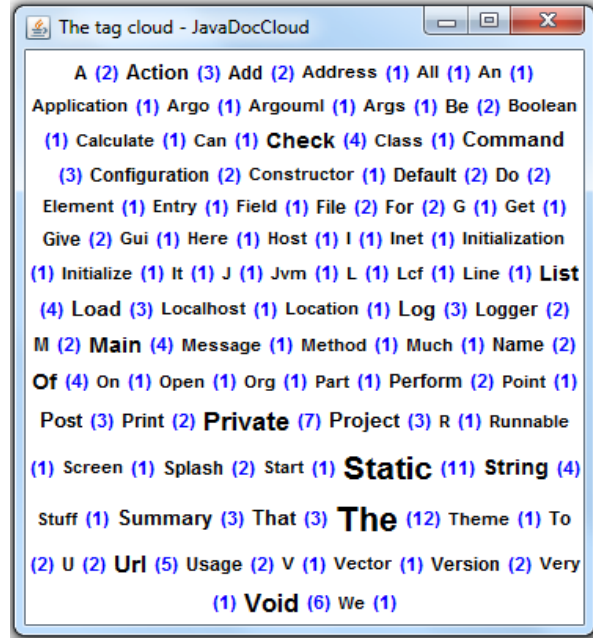

Fig. 3 The tag cloud with tag frequency for the main class of ArgoUML software

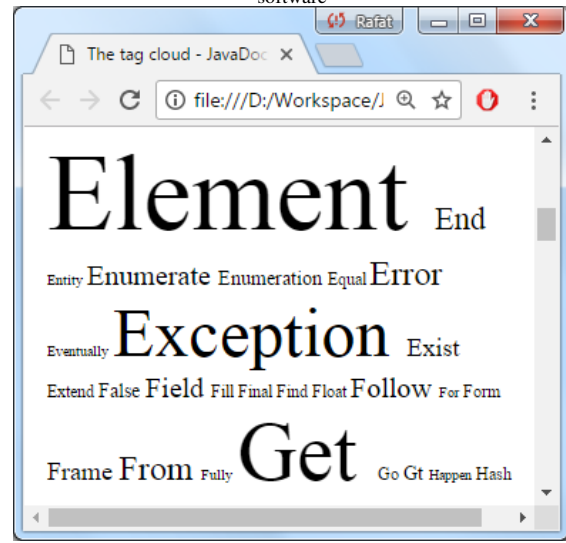

Fig. 4 The extracted tag cloud from JavaDoc of NanoXML software

The most common and uncommon tags across the Javadoc documents of NanoXML software is presented in Table II.

TABLE II
TAGS MINED FROM NANOXML JAVADOC

| The most common tags | | The uncommon tags (e.g.) | |
|---|---|---|---|
| Tag | Frequency | Tag | Frequency |
| Java | 140 | Validate | 1 |
| The | 136 | Serializable | 3 |
| String | 117 | Override | 1 |
| Property | 97 | Note | 1 |
| Lang | 91 | Jar | 2 |
| The number of tags = 224 tags | | | |

Figure 5 shows the tag cloud of NanoXML JavaDoc documents. In Figure 5, the tag cloud includes the tag frequency. In JavaDocCloud approach the tag frequency uses as an indicator of the tag frequency through the software documents. For a lack of research evaluating tag clouds in the software engineering field, there was trouble in evaluating the JavaDocCloud method.

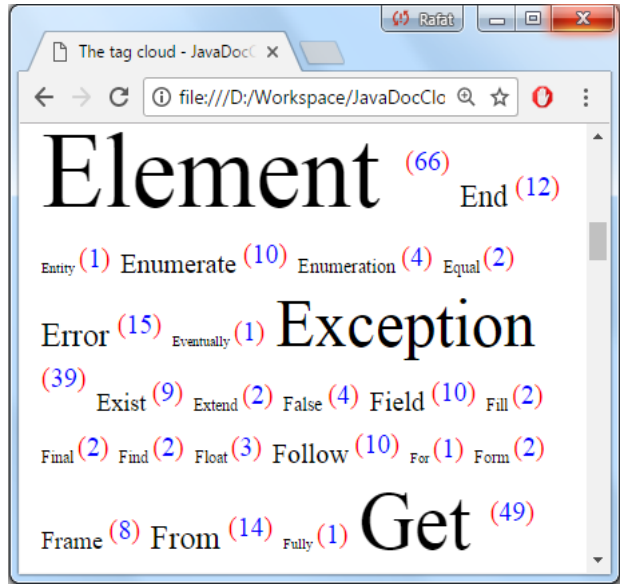

Fig. 5 The tag cloud with tag frequency for the NanoXML JavaDoc

The threat to the validity of JavaDocCloud approach is that current work considers only one type of software documents which is JavaDoc. The Wordnet dictionary maybe not reliable in all cases to identify word stems or roots. Moreover, when software developer uses mix words (such as: GeTSettingS) in the JavaDoc documents the camel-case splitting algorithm can't handle such word.

## V. CONCLUSION

This paper presented a new technique to visualize software documents as a tag cloud. The key idea of this approach is to help software experts to understand the legacy software documents via tag cloud technique. The novelty of JavaDocCloud approach is that it exploits the Javadoc of software system to build an efficient tag cloud. The approach has applied on NanoXML case study. The resulted tag cloud shows the most frequent tags and the uncommon tags. Regarding future work, JavaDocCloud approach plans to extend the current approach to include other types of software artifacts such as: design documents.

REFERENCES

[1] J. Emerson, Tag Clouds in Software Visualisation, MSc Thesis, University of Canterbury, 2014.
[2] J. Emerson, N. Churcher, A. Cockburn, "Tag Clouds for Software and Information Visualisation", 14th Annual ACM SIGCHI NZ Conference on Computer-Human Interaction, Christchurch, New Zealand, November 15-16, pp. 1-4, 2013.
[3] J. Emerson, N. Churcher, C. Deaker, "From Toy to Tool: Extending Tag Clouds for Software and Information Visualisation", 22nd Australian Software Engineering Conference, Melbourne, Australia, June 4-7, pp. 155-164, 2013.
[4] R. Al-Msie'deen, "Tag Clouds for the Object-Oriented Source Code Visualization," Engineering, Technology & Applied Science Research, vol. 9, no. 3, pp. 4243–4248, 2019.
[5] D. Kramer, "API documentation from source code comments: a case study of Javadoc," In Proceedings of the 17th annual international conference on Computer documentation, SIGDOC '99, pp. 147–153, 1999.
[6] C. Deaker, L. Pettigrew, N. Churcher, and W. Irwin, "Software visualisation with tag clouds," in ASWEC 2010 Industry Track Proceedings, J. Hosking and B. Long, Eds., Auckland, New Zealand, pp. 129–133, 2010.
[7] R. Cottrell, B. Goyette, R. Holmes, R. J. Walker, J. Denzinger, "Compare and Contrast: Visual Exploration of Source Code Examples", 5th IEEE International Workshop on Visualizing Software for Understanding and Analysis, Edmonton, Canada, September 25-26, pp. 1-4, 2009.
[8] C. Anslow, J. Noble, S. Marshall, E. D. Tempero, "Visualizing the Word Structure of Java Class Names", in Companion to the 23rd Annual ACM Sigplan Conference on Object-Oriented Programming, Systems, Languages, and Applications, Nashville, USA, Octomber 13-19, 2008.
[9] M. Stocker, https://misto.ch/2011/09/19/tag-cloud-visualization-for-source-code/, August 25, 2019.
[10] J. Martinez, T. Ziadi, T. F. Bissyande, J. Klein, Y. L. Traon, "Name Suggestions During Feature Identification: The Variclouds Approach", 20th International Systems and Software Product Line Conference, Beijing, China, September 16-23, 2016.
[11] ArgoUML Javadocs-0.20: http://argouml-stats.tigris.org/, August 25, 2019.
[12] R. Al-Msie'deen, M. Huchard, A. Seriai, C. Urtado, S. Vauttier, "Automatic documentation of [mined] feature implementations from source code elements and use-case diagrams with the REVPLINE approach", International Journal of Software Engineering and Knowledge Engineering, Vol. 24, No. 10, pp. 1413–1438, 2014.
[13] R. Al-Msie'deen, A. D. Seriai, M. Huchard, C. Urtado, S. Vauttier, "Documenting the mined feature implementations from the object-oriented source code of a collection of software product variants", 26th International Conference on Software Engineering and Knowledge Engineering, Knowledge Systems, Vancouver, Canada, July 1-July 3, 2014.
[14] G. A. Miller, "Wordnet: A lexical database for English", Communications of the ACM, Vol. 38, No. 11, pp. 39–41, 1995.
[15] WordNet: https://wordnet.princeton.edu, August 25, 2019.
[16] ArgoUML Javadocs: http://argouml-stats.tigris.org/nonav/javadocs/javadocs-0.20/, August 25, 2019.
[17] NanoXML: http://nanoxml.sourceforge.net/orig/index.html, August 25, 2019.
[18] R. Al-Msie'deen, https://sites.google.com/site/ralmsideen/tools, August 25, 2019.